# A framework for designing experimental tasks in contemporary physics lab courses

**Simon Z Lahme[1,a], Pekka Pirinen[2], Lucija Rončević[3], Antti Lehtinen[2,4], Ana Sušac[3], Andreas Müller[5] and Pascal Klein[1]**

[1] Faculty of Physics, Physics Education Research, University of Göttingen, Friedrich-Hund-Platz 1, 37077, Göttingen, Germany
[2] Department of Physics, University of Jyväskylä, P.O. Box 35, 40014 University of Jyväskylä, Finland
[3] Department of Applied Physics, Faculty of Electrical Engineering and Computing, University of Zagreb, Unska 3, 10000, Zagreb, Croatia
[4] Department of Teacher Education, University of Jyväskylä, P.O. Box 35, 40014 University of Jyväskylä, Finland
[5] Faculty of Sciences, Department of Physics, University of Geneva, Boulevard du Pont d'Arve 40, 1211, Genève, Switzerland

[a] email: simon.lahme@uni-goettingen.de

**Abstract**. While lab courses are an integral part of studying physics aiming at a huge variety of learning objectives, research has shown that typical lab courses do not reach all the desired goals. While diverse approaches by lab instructors and researchers try to increase the effectiveness of lab courses, experimental tasks remain the core of any lab course. To keep an overview of these developments and to give instructors (and researchers) a guideline for their own professional efforts at hand, we introduce a research-informed framework for designing experimental tasks in contemporary physics lab courses. In addition, we demonstrate within the scope of the EU-co-funded *DigiPhysLab*-project how the framework can be used to characterize existing or develop new high-quality experimental tasks for physics lab courses.

## 1. The outstanding role of experimental tasks in university physics education

Besides lectures and exercises, lab courses have a long tradition in studying physics both at high schools and universities. As described in [1], there has been a significant shift from lecture-based physics education to hands-on student lab activities throughout the late 1800s and early 1900s with the aim of "emphasiz[ing] 'the development of habits of scientific thought' and 'the method by which science obtains its results' rather than 'more or less scattered facts and theories' taught in such a way that they could only be committed to memory" (p.53). This focus on teaching the scientific "practice of inducing principles from data" (p.54) remained until today [1].

Nowadays, lab courses are undoubtedly an integral part of physics education with a significant percentage of university physics studies programs (e.g., in Germany 11% - 22% of the workload in Bachelor study programs are officially recommended and most study programs follow this directive [2]). This can also be seen in the increasing number of recommendations for learning objectives to be reached in such physics lab courses (e.g., Refs. [3,4] for the United States; Refs. [5,6] for Germany; Ref. [7] for six European countries).

While in theory, there is broad consensus about the importance of these learning objectives for physics lab courses, research has distinctly shown that typical lab courses do not reach the desired goals. For example, they do not enhance the students' concept- or calculus-based factual knowledge [8], do not support expert-like views and attitudes towards experimental physics [9], and do not meet the students' interests [10]. They also provide many opportunities for rather unmeaningful activities like manipulating the setup and taking measurement data instead of more meaningful activities like conducting a quick evaluation or discussing ideas with other students and instructors [11], so typical lab courses rarely provide engaging learning opportunities for critical thinking [8].

Over the last decades, different approaches have been pursued to increase the effectiveness of physics lab courses e.g., by stressing the principle of *open inquiry-based learning* [8,12,13], meeting the addressees' specific needs (e.g., for medicine students [14]), following the principle of *cognitive apprenticeship* [15], or integrating modern digital technologies like *smartphones* [16,17] into lab courses and physics education.

In this dynamic field of innovation and research, one characteristic of physics lab courses remains crucial: in the centrum of each lab course are experiments that are carried out by the students more or less guided by (written) task instructions and usually an instructor. In other words: while the above outlined findings and developments have a significant impact on the conception and design of experimental tasks, these tasks are (together with the instructor's guidance) still *the* main learning opportunity to reach the learning objectives of lab courses. They remain the focal point of contemporary and effective lab courses, so these innovative concepts still follow the approach of task-based learning [18] which considers tasks as the centrum of a constructivist learning process.

Here, we use the term *experimental task* in distinction to others like *task*, *(task) instructions*, *experiment*, *experiment guide(-lines)*, or *labwork* to denote the full conceptualization of a time-limited, task-based learning environment, but with a minor focus on the interaction between students and instructor. Thus, *experimental tasks*, as understood here, consist of an idea/concept of an experiment- and a task-based learning environment with materials like task instruction sheets, lab equipment, etc.

With this dominant role of experimental tasks for the students' learning processes in a lab course, two research needs become self-evident. On the one hand, research is needed on how to design high-quality experimental tasks for physics lab courses (as it is already done widely, cf. the cited literature above). On the other hand, those research findings need to be integrated and communicated to the target group of instructors who are responsible for the design and conceptualization of lab courses.

In this paper, we contribute to the second demand. As the development and research related to the design of high-quality experimental tasks are accumulating nationally and internationally (cf. first demand), an overview of design principles for experimental tasks in physics lab courses would be beneficial for researchers and lab instructors to reflect and control their respective professional efforts (cf. second demand). Hence, we process the state of research and communicate findings to the field of actors by answering the question: *Which design principles can be considered to design new or characterize existing experimental tasks for contemporary physics lab courses?*

For this purpose, we present in Sec. 3, based on a literature review in Sec. 2, a framework with design principles for experimental tasks in contemporary physics lab courses. In Sec. 4, we outline how the framework can be used to characterize existing and develop new experimental tasks within and outside the related, EU-co-founded *DigiPhysLab*-project (Developing Digital Physics Lab Work for Distance Learning; cf. Ref. [19]). Sec. 5 finally contains a summary and future steps to be taken.

## 2. State of research regarding the design and/or taxonomy of experimental tasks

In preparation for the development of our framework, we reviewed the literature regarding the design and/or taxonomy of experimental tasks. On the one hand, the review revealed several frameworks and taxonomies regarding aspects to be considered during the design or characterization of experimental tasks which are summarized in Sec. 2.1 and will be integrated into our framework later (cf. Sec. 3). On the other hand, two general approaches to designing learning environments were found which are described in Sec. 2.2 and will provide the structural basis of our framework.

*2.1. Aspects for the design and characterization of experimental tasks found in the literature*
Various literature sources regarding the design and characterization of experimental tasks can be related to desired learning objectives. There are several explicit catalogs with learning objectives for lab courses created either normatively [3,5] or empirically [4,6,7,20]. Some take the form of short lists of educational objectives [21,22], while other frameworks address specific learning objectives in detail, e.g., the acquisition of digital competencies [23,24] or the growth of adequate conceptions regarding experimental physics [9].

The compilation of learning objectives is closely related to thoughts about possible students' activities during lab courses [11,20,21,25-27] as well as the role of collaboration [21,22,26,28] and instructors' guidance [20,26,28] during those activities. Accordingly, the openness/closure of the activities and therefore the experimental tasks is discussed [20,26,29,30]. This is linked to the underlying learning theory approach of the experimental task ranging from more guided approaches over inquiry-based learning approaches toward very open undergraduate research projects [8,12,22,28,29,31]. The latter is also connected with the logical function of the experimental task in the overall learning process [20,22,26,29,32].

Further aspects discussed in the literature refer to the levels of difficulty [33], the nature of factual encounter and artificiality of data collection and used equipment [21,26], the areas of digitalization in lab courses [34] as well as affective, metacognitive, and social dimensions of experimenting [22,28]. Additionally, the literature review reveals ideas for the actual implementation of experimental tasks, e.g., regarding the delivery of the task to the students [20], the consideration of supportive materials [32], the students' record and assessment [20,32], or boundary conditions to be noted like costs for equipment [21]. Finally, one can also apply general criteria for high-quality physics tasks to the design of experimental tasks, e.g., linguistic simplicity, brevity, conciseness, correctness, etc. [18].

*2.2. Approaches for the design of a learning environment as the structural basis for the framework*
For the design of a learning environment (e.g., a task, a lesson, …) in general and therefore applicable for the design of experimental tasks, two well-known models/approaches can be found in the literature: The first one, the *Model of Educational Reconstruction* (abbr. as MER, [35]), was originally invented for the overall design of science learning environments especially at schools. Its core idea is that instructors should follow three steps while designing learning environments: First, they *clarify and analyze the science content* they want to teach (e.g., the relevance of the content, its subject systematic structure, or the learning goals). Second, they *investigate into their students' perspectives regarding the selected science content* (e.g., the students' interests, preconceptions, or prior knowledge); this step can of course be done in reciprocity with the first step. Finally, findings are used to iteratively integrate the clarification and analysis of the science content and the investigation into students' perspectives to *design and evaluate the learning environment* [35], which in our case is an experimental task.

The second approach, *Action Research* (abbr. as AR, [36]), is a research design for iteratively solving practical problems, especially by educators with the aim of improving their educational practice by understanding, evaluating, and changing. It describes an ongoing process of innovation consisting of several linked cycles of four steps: *plan*, *act*, *observe*, and *reflect*. "*From the point of view of teachers and teaching, it involves deciding on a particular focus for research, planning to implement an activity, series of activities, or other interventions, implementing these activities, observing the outcomes, reflecting on what has happened and then planning a further series of activities if necessary*" [36, p.7]. This process leads to a high-quality action/intervention as a solution for the initial (educational) problem. In the case of our framework, this is a high-quality experimental task ready to be implemented into university physics teaching, so that the target group of students achieves the desired learning goals.

## 3. The framework for designing experimental tasks in contemporary physics lab courses
In the literature review in Sec. 2, we listed several existing frameworks/taxonomies related to the design of experimental tasks. But so far, to our knowledge, there is no uniform framework that focuses on the design process itself and integrates already existing findings and frameworks. As we already discussed in Sec. 1, such an overview of design principles for experimental tasks in physics lab courses would be beneficial for researchers and lab instructors to reflect on and control their respective professional efforts and would narrow the gap between existing research findings and teaching practices.

Thus, we developed a framework (cf. Figure 1) for designing experimental tasks in contemporary physics lab courses providing a list of six design principles. They are arranged in the pattern of a workflow guiding lab instructors in their development of experimental tasks for physics lab courses.

As explained below, the overall structure is based on the two approaches MER and AR described in Sec. 2.2. Design principles 1 to 3 can be linked to the MER, and design principles 3 to 5 can be linked to the AR approach (design principle 3 is bridging both approaches); the last design principle is neither linked to the MER nor the AR approach. Furthermore, each design principle comes along with a list of research-informed categories that explicate what needs to be considered in each step.

A prior version of the framework was already sketched in Ref. [19] but it has been modified significantly as it now provides real design principles and a workflow for designing experimental tasks.

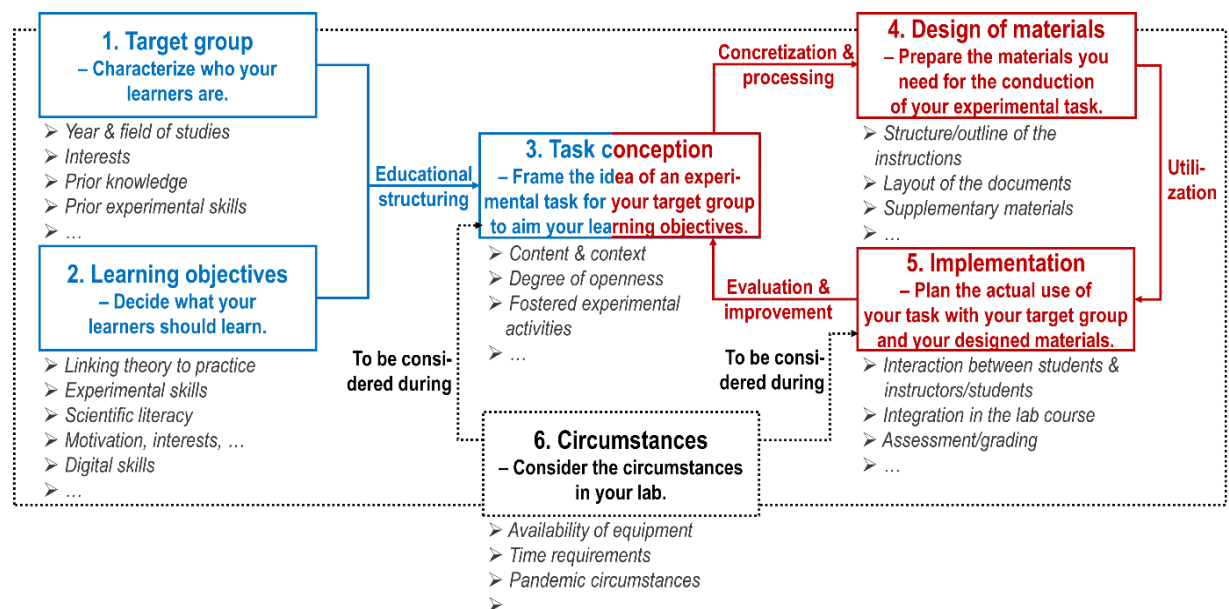

**Figure 1.** Framework for designing experimental tasks in contemporary physics lab courses. The left part is based on the Model of Educational Reconstruction (MER, in blue), and the right part is based on the Action Research approach (AR, in red). Everything is affected by the circumstances (dotted). The framework provides research-informed categories for each design principle (in italics).

The six design principles in our framework are:

**1. Target group – Characterize who your learners are.** Our first design principle is about analyzing the target group of the experimental task to be developed. It largely corresponds to the *investigation into students' perspectives* in the MER and is the starting point for an addressee-oriented task design. To characterize the target group, it contains the categories *year of study* and *field of study* as well as *expected prior knowledge* and *expected prior experimental skills* addressing the preconditions of the target group that need to be considered to design an addressee-specific experimental task (based on Refs. [14,18]). For the same reason, it is e.g., also relevant to be aware of the *students' interests* (based on Ref. [10]) and *attitudes* (based on Ref. [18]).

**2. Learning objectives – Decide what your learners should learn.** Our second design principle is about stating the learning objectives of the experimental task to be developed. It accentuates the *clarification and analysis of the science content* in the MER with an emphasis on the learning objectives. However, this emphasis is not a restriction of that step in the MER as e.g., analyzing the relevance of the content or its subject systematic structure are necessary steps for clearly stating learning objectives.

This design principle is directly linked to the catalogs of learning objectives mentioned in Sec. 2.1 and provides a list of objectives to be reached. It is based mostly on work described in Ref. [7] leading to the four categories of learning objectives *linking theory to practice*, *learning experimental skills*, *getting to know the methods of scientific thinking,* and *fostering motivation*, *personal development, and social competency*. Further literature about learning objectives e.g., Refs. [3-6,9,20-22,28] is integrated here as subcategories. Additionally, there are two categories of learning objectives *acquiring digital competencies* to integrate ideas from Refs. [23] and [24] and *acquiring writing and presentation competencies* based on Refs. [4] and [6].

**3. Task conception – Frame the idea of an experimental task for your target group to aim your learning objectives.** Our third design principle refers to the overall conceptualization of the experimental task based on the analysis of the target group and the decision on the learning objectives. Thus, a process of *educational structuring* is required here. On the one hand, this is comparable to the third step in the MER; just a slight difference is that in our framework this step does not include the complete development of all learning materials (e.g., task instructions) but focuses on a preliminary task conceptualization (e.g., regarding scope, requirement level, or degree of openness). On the other hand, the third design principle is also the starting point for the AR approach in our framework as one needs to come up with an initial idea and concept of the experimental task; so, the task conception is related to the *planning* step in the AR approach, too.

Here, we want to note that third step can also lead to the decision that an experimental task does not fit with the needs of the target group and the desired learning objectives, i.e., that a lab course is not suitable for reaching the desired learning objectives with the specific target group. In that case, developing an experimental task is not conclusive, so further design principles can be disregarded.

The third design principle contains all ideas from the literature review corresponding to the overall design and conceptualization of experimental tasks. Thus, there is the category *logical function of the experiment* (based on Refs. [20,22,26,29,32,34]) to address the different purposes of using an experimental task and its integration in the whole learning process. Another category is the *degree of openness* referring to the underlying learning theory approach of experimental tasks and the opportunities to vary the degrees of guidance and openness (based on Refs. [8,12,18,20,22,26,28,29,31,32]). Related to that, there is also the category *level of difficulty* (based on Refs. [18,28,33]) as one can vary the difficulty of the experimental task considering the target group, learning objectives, and openness of the task. Additionally, this design principle contains a list of *focused experimental activities* that can be stimulated among the students with the experimental task (based on Refs. [11,20,21,25-27,34]) as well as the categories *learner-object-relationship* and *mode of data collection* to include the aspects nature of factual encounter and artificiality of data collection and used equipment (based on Refs. [21,26,34]). Further, more descriptive categories are the *topic/content* and *context* of the task or a list of *necessary digital technologies* and *necessary further equipment*.

**4. Design of materials – Prepare the materials you need for the conduction of your experimental task.** Our fourth design principle is specifically linked to the design of the learning materials for the experimental task, e.g., the task instructions, so it requires a concretization and processing step to design the materials based on the *task conception*. Thus, the fourth design principle is still part of the *planning* step in the AR approach but from our point of view, it is reasonable to distinguish between *task conception* and *design of materials* because the design takes place on two different levels which should mostly be done in a fixed order: Before thinking of actual instructions for the students, specific learning materials, etc., one should first determine the overall experimental task by answering questions like: *To which content is this task related? What should my learners learn? Which equipment should be used? How open should the task be?* Only when these questions have been answered, a meaningful design of the learning materials can take place. Now different questions need to be considered: *How much theoretical background needs to be provided so that my students understand the physics behind the experimental task? How do the task instructions need to be structured and formulated so that my students can reach my learning objectives? How do the task instructions need to be formulated so that my students can meaningfully use the selected equipment while the task is as open as intended?*

Hence, there are categories like *structure/outline of the instructions* referring to the logical order of the instructive task documents, the *layout of the documents* referring to the visual appearance of the task documents, and the *choice of words and content* such as the text length, the linguistic simplicity, or the availability of meaningful graphical representations which can influence the task difficulty [18]. Additionally, one can think of *supplementary materials* (based on Ref. [32]) provided to the students during the learning process and the *correctness of the materials* (based on Ref. [18]) since all given information must be correct.

**5. Implementation – Plan the actual use of your experimental task with your target group and your designed materials.** The fifth design principle of our framework addresses the implementation of the designed experimental tasks in line with the *action* step in the AR approach. Here, one needs to think of the actual utilization of the designed materials with the target group, so new aspects need to be considered e.g., the interaction among the students or the instructors' guidance. Thus, there are the categories *social form of learning* to address the degree of collaboration among the students (based on Refs. [21,22,26,28,34]) and *interaction between students and instructor* (based on Refs. [20,26,28,34]) that is related to the guidance/openness of the task but focuses more on mode and circumstances of interaction. Additional categories are *delivery of the task* (based on Ref. [20]) to describe the mode of communication of the task instructions and *integration in the course* (based on Refs. [18,26]) to address the organizational realization of the experimental process. Further categories are *assessment/grading* of students' learning progress (based on Ref. [20,32,34]) and *feedback* provided for students to reflect on their learning process and for instructors to evaluate the experimental task (based on Ref. [34]).

Based on the experiences with the actual implementation, an *evaluation and improvement* of the experimental task can follow. Then, one continues with the third design principle *task conception* and starts with a re-design of the experimental task. Here, the other steps of the AR approach, *observing* and *reflecting* as well as the next cycle of this iterative process are located in our framework: Based on the observations of the implementation of the experimental task one can reflect on the design also concerning target group, intended learning objectives, and given circumstances. The reflection can either lead to the decision that the experimental task meets the needs (and therefore solves the initial problem) or that it needs to be modified. In the latter case, the improvement and re-pilot of the experimental task follow, so a new cycle of planning, acting, overserving, and reflecting is performed.

**6. Circumstances – Consider the circumstances in your lab.** The last design principle acknowledges the variety of circumstances that obviously affect the implementation of the experimental task and, therefore, also need to be considered during the task conception (cf. design principles 3 and 5). It contains organizational categories like *time requirements*, *availability of equipment*, and *costs for the equipment* (based on Ref. [21]). Furthermore, there is the category *pandemic circumstances* referring to special needs of distance learning and hygienic measures. Finally, there is the category *limitations for implementation* where one can reflect any relevant limitations, e.g., the availability of cited literature in only one language or the usability of specific software only on one operating system. The sixth design principle is not explicitly part of either the MER or the AR approach but is an integral part of the actual (educational) problem in the AR approach as the starting point for the action/intervention process.

All in all, we have provided six fundamental design principles linked to each other to form a coherent workflow for designing experimental tasks for contemporary physics lab courses which are based on the MER and the AR approach. We want to argue that both approaches are needed for the theoretical basis of our framework because even though the MER includes the iterative design of the learning environment (while integrating students' perspectives and clarifying/analyzing the science content) it does not stress the steps of *implementation*, *evaluation*, and *improvement* of the designed learning environment as much as it is done by the AR approach with the steps *acting*, *observing*, and *reflecting*.

## 4. The use of the framework exemplified in the scope of the *DigiPhysLab*-project and beyond

The above-presented framework provides a direct answer to our introductory question. It can be utilized in two ways: On the one hand, the framework can be used to characterize existing tasks and to showcase their similarities and differences which supports the reflection of one's own lab tasks. On the other hand, the framework can serve as an inspiration and guidance for researchers and lab instructors to develop new experimental tasks for contemporary and effective physics labs. These two use cases will be outlined in more detail in the following section, first in the scope of the EU-co-funded *DigiPhysLab*-project and after that in a project-independent, more generalized manner.

*4.1. The EU-co-funded project DigiPhysLab*
The framework is one of the three main intellectual outputs of the *DigiPhysLab*-project (Developing Digital Physics Lab Work for Distance Learning), an education and research project among the three participating universities in Göttingen (Germany), Jyväskylä (Finland), and Zagreb (Croatia) co-funded by the Erasmus+ program of the European Union (running time 03/2021-02/2023). As outlined in Ref. [19], the primary goal is the development and evaluation of 15 competence-centered, high-quality experimental tasks for university physics education suitable both for distance learning scenarios (e.g., during the COVID-19 pandemic) and on-campus teaching. To enable students to conduct hands-on experiments e.g., even in the students' private homes, we utilize (alongside household items) modern digital technologies like smartphones for data collection and data processing software for data analysis. For each experimental task, we prepare explicit task instructions ready to be used with students as well as additional instructions with background information only for instructors. The tasks are piloted with students in our three faculties and improved based on the evaluation findings.

*4.2. Framework-based task development and characterization in the DigiPhysLab-project*
In the *DigiPhysLab*-project, the whole workflow is accompanied by the presented framework in three ways. First, we use the framework for the development of our own 15 experimental tasks as we follow the same workflow and design principles as described in the framework. There, we must consider that our tasks to be developed should both be piloted (and implemented) in our faculties and be published as *Open Educational Resources* for other lab instructors simultaneously. Hence, we treat the six design principles in two different ways: We decide for each of our tasks everything related to the design principles *learning objectives*, *task conception*, and *design of materials*. Therefore, aspects of our experimental tasks related to these three design principles are predetermined by us in the students' versions of our task documents. These decisions are immanently linked to our idea of the experimental tasks and need to be considered during the implementation in our faculties and by other instructors who want to use our tasks for their own lab courses. Oppositely, aspects of our experimental tasks related to the design principles *target group*, *implementation*, and *circumstances* are dependent on the actual usage of our task documents and are therefore not predetermined by us; instead, we only provide suggestions related to these three design principles in the instructors' versions of our task documents based on our experiences from the task evaluation and the circumstances in our three faculties. Lab instructors can use these suggestions as a guide for their own implementation of our experimental tasks and for their modifications to adapt our tasks to their local conditions.

The second purpose of using the framework in the *DigiPhysLab*-project is to characterize our 15 experimental tasks after the development process has ended. By this, we can clearly and briefly describe the characteristics of our tasks and reflect on their diversity. Table 1 outlines what this framework-based task characterization can look like exemplified by the experimental task *Slamming door* (cf. task documents on our project website, www.jyu.fi/digiphyslab). In that task, students replicate an experiment described in Ref. [37] in which they investigate the occurring frictional effects of a door when it is slammed shut. They collect data with the acceleration or gyroscope sensor of their smartphone and statistically compare the quality and validity of different frictional models.

Third, the framework is reflected in the evaluation instrument (cf. Ref. [38]) developed in the *DigiPhysLab*-project to evaluate our experimental tasks during the pilots with students in our three faculties. Especially the categories *fostered experimental activities* and *openness of the task* and items about the adequacy/quality of the materials for the target group are part of the instrument.

All in all, the framework is the basis for the whole workflow in the *DigiPhysLab*-project and particularly serves as a tool to showcase the scope of the students' and instructors' task documents.

**Table 1.** Outlined usage of the framework for the characterization of an experimental task.

| Framework – *Excerpt* | Characterization of the task *Slamming Door* |
|---|---|
| 1. Year/field of studies | 1st year physics major (and teacher training) students |
| 2. Linking theory to practice | Describing a real rotatory movement with physical models, understanding the importance of model parameters for the model validity |
| 3. Content/context | Mechanics: real rotary motion of a slamming door with friction |
| 3. Degree of openness | Goal & research question given, slightly guided data collection & analysis |
| 3. Fostered experimental activities | Collecting data with a smartphone, analyzing, and fitting data, comparing the quality/validity of different models to fit measurement data |
| 4. Structure/outline of the instructions | Brief motivation and introduction into theory, task with vague instructions for the experiment, guiding questions, supportive materials & literature |
| 5. Interaction | Students collaborate in groups of 2-4, instructor's guidance on demand |
| 5. Assessment | E.g., a lab report, a scientific poster, … |
| 6. Equipment | Smartphone, everyday door, tape, folding rule, computer for data analysis |
| 6. Time requirements | Minimum 2 h, better would be 3-4 h (assessment excluded) |

*4.3. Potential purposes of using the framework beyond the DigiPhysLab-project*

Outside the *DigiPhysLab*-project, our framework can largely serve the same purposes of use both for lab instructors and physics education researchers to reflect and control their professional efforts. First, the framework can be used by instructors (and researchers) to develop new high-quality experimental tasks as it describes a workflow and principles of how to design experimental tasks for contemporary physics lab courses. Second, the framework can be used to characterize already existing experimental tasks. By this, instructors can reflect on and evaluate the quality and diversity of the tasks in their own lab courses and researchers have a tool at hand to systematically review different lab concepts. Third, the framework can also be used to get an overview of design principles and adjustment screws to be considered for designing experimental tasks. Thereby, instructors become aware of the huge variety of opportunities to design new or modify existing experimental tasks and physics education researchers can map their research to the framework and can identify new research needs and questions. Fourth, the framework could also be beneficial for physics teachers in (high) schools and colleges for the same reasons as mentioned before. Overall, we hope that our framework contributes to the development of innovative, effective, high-quality experimental tasks for physics lab courses and narrows the gap between research findings and teaching practices in laboratory physics education.

**5. Summary and outlook**

We presented a research-informed framework about design principles for experimental tasks in contemporary physics lab courses. It is based on the *Model of Educational Reconstruction* and the *Action Research* approach and consists of six design principles that need to be considered during the design process (*target group, learning objectives, task conception, design of materials, implementation,* and *circumstances*), which are completed by multiple aspects from the state of research. The framework can serve lab instructors and physics education researchers as a tool for the development and characterization of experimental tasks as well as the reflection of the variety of aspects related to the design of and research about experimental tasks in lab courses.

In the future, several further steps can be taken to improve the framework and to advance its purposes of use. First, further literature can be integrated into the already existing framework (which functions as a backbone) and each aspect of the design principles can be provided with an operationalizing vocabulary list that allows a uniform task characterization (to advance the backbone to a full corpus). Second, the framework should be validated with lab instructors and physics education researchers to ensure usability and practicability for its addressees. Finally, experimental tasks of different lab courses can be characterized with the framework to create a review of labwork in different institutions/countries.

**Author contributions**[*]
**P Klein**: Conceptualization (supporting), Funding Acquisition (supporting), Supervision (equal), Writing – review & editing (equal). **S Z Lahme**: Conceptualization (lead), Visualization, Writing – original draft, Writing – review & editing (equal). **A Lehtinen**: Conceptualization (supporting), Funding acquisition (lead), Project administration, Supervision (equal), Writing – review & editing (equal). **A Müller**: Supervision (equal), Writing – review & editing (equal). **P Pirinen**: Conceptualization (supporting), Writing – review & editing (equal). **L Rončević**: Conceptualization (supporting), Writing – review & editing (equal). **A Sušac**: Conceptualization (supporting), Funding Acquisition (supporting), Supervision (equal), Writing – review & editing (equal).

[*] According to CRediT (CRediT Contributor Roles Taxonomy), https://credit.niso.org/

**Funding**
We are grateful for the financial support by the Erasmus+ program of the European Union (G.A.-No.: 2020-1-FI01-KA226-HE-092531).